\documentclass[preprint2]{aastex631}

\usepackage{ulem}

\definecolor{HTcolor}{RGB}{0,128,0}
\definecolor{draftcolor}{RGB}{192,0,192}

\received{...}
\revised{...}
\accepted{...}
\submitjournal{ApJ}

\begin{document}

\title{The relation between Simulated Multiwavelength Blazar Variability and Stochastic Fluctuations}


\author[0000-0002-4024-3280]{Hannes Thiersen}
\affiliation{Centre for Space Research, North-West University, Potchefstroom, 2520, South Africa}

\author[0000-0001-5801-3945]{Michael Zacharias}
\affiliation{Landessternwarte, Universit\"at Heidelberg, K\"onigstuhl 12, 69117 Heidelberg, Germany}
\affiliation{Centre for Space Research, North-West University, Potchefstroom, 2520, South Africa}

\author[0000-0002-8434-5692]{Markus B\"ottcher}
\affiliation{Centre for Space Research, North-West University, Potchefstroom, 2520, South Africa}


\begin{abstract}

Blazars exhibit multiwavelength variability, a phenomenon whose underlying
mechanisms remain elusive. This study investigates the origin of such variability
through leptonic blazar emission simulations, focusing on stochastic fluctuations
in environmental parameters. By analyzing the spectral indices of the power
spectral densities of the variability, we assess their relationship with the
underlying fluctuations. Our findings reveal that the variability spectral indices
remain almost independent of the variations responsible for their emergence. This suggests
a complex interplay of factors contributing to the observed multiwavelength
variability in blazars.

\end{abstract}

\keywords{--}

\section{Introduction}
\label{sec:introduction}

Blazars are a subclass of active galactic nuclei known to be some of the brightest persistent multi-wavelength sources in the sky \citep{urry_and_padovani_1995,abdollahi_2020}.
Blazars possess jets of relativistic particles propagating along their central region's rotation axis which is closely aligned to our line of sight.
The relativistic bulk motion of the emission region leads to highly Doppler-boosted observed emission.
Radio up to UV/X-ray radiation is produced by leptonic synchrotron emission, while the mechanisms producing the high-energy spectral component (X-ray through $\gamma$-rays, in some cases up to TeV energies) are still up for debate.
Some candidate mechanisms include inverse Compton scattering \citep{begelman_and_sikora_1987,maraschi_1992,dermer_and_schlickeiser_1993,sikora_1994,ghisellini_2010,böttcher_2013}, proton synchrotron radiation, and photo-pion production and subsequent decay of pions and muons, initiating electromagnetic cascades \citep{mannheim_and_bierman_1992,mannheim_1993,mücke_and_protheroe_2000,böttcher_2013}.

Blazar emission exhibits unpredictable variability across the entire electromagnetic spectrum.
This variability is observed on both short (minutes up to days) and long (weeks up to years) time scales \citep[e.g.,][]{albert_2007,aharonian_2007,ackermann_2016,arlen_2013,chatterjee_2021}.
Multiwavelength blazar observation campaigns \citep[e.g.,][]{nilsson_2018,abdollahi_2020,penil_2024,MAGIC_2024} elucidated the phenomenology of blazar variability in some detail.
However, the exact cause(s) of this phenomenon still elude(s) the scientific community.

One of the characteristics routinely found in long-term observations of blazar variability as well as Galactic X-ray binaries is either simple or broken power-law spectra in the power spectral densities (PSDs) of their light curves \citep{HESS_2017,goyal_2020,bhatta_and_dhital_2020,goyal_2022}.
Therefore it is plausible that some type of stochastic process is at work producing the multiwavelength variability such that emission power and temporal frequency are anti-correlated, i.e. $P(f) \propto f^{-\alpha}$, also commonly referred to as coloured noise.
Furthermore, some works quantifying the power law indices have found that these indices do not necessarily agree between different wavebands \citep[e.g.][]{goyal_2022}.

Various models have been developed that exhibit colored noise variability, but are often constrained by the duration of particular events, rendering them effective only temporarily.
Very few models attempt to explain continuous sustained variability over the large range of observed variability time scales \citep[e.g.,][]{finke_and_becker_2015}.
Recent papers, including \cite{tavecchio_2020, adams_2022, brill_2022}, employ stochastic differential equations to model high-energy blazar variability with some success, adding to the growing evidence for the role of stochastic processes in these phenomena.

This work attempts to understand how variability can arise assuming some self-sustaining process manipulates key parameters in the emitting region in a stochastic manner.
Details of the model and methodology are explained in Section~\ref{sec:model_setup}.
Section~\ref{sec:results} presents the main results of the paper followed by a discussion and conclusions in Section \ref{sec:discussion_and_conclusions}.

\section{Model Setup}
\label{sec:model_setup}

\subsection{Leptonic one-zone Emission Model}
\label{sub:leptonic_one-zone_emission_model}

We use a leptonic one-zone blazar emission model to simulate multi-wavelength blazar variability \citep{diltz_2014, zacharias_2017, zacharias_2022, thiersen_2022, zacharias_2023}.
The model assumes that all emission is produced by a single homogeneous volume/zone located close to the central region, containing isotropically distributed relativistic leptons (electrons and possibly positrons).
Due to the relativistic nature of the leptons they produce non-thermal emission via synchrotron (radio up to UV and/or X-ray radiation) and inverse-Compton radiation (X-ray up to very high-energy [VHE] $\gamma$-rays).
Furthermore, the blob travels at relativistic speed with bulk Lorentz factor $\Gamma$ with respect to the central region.
The observer's line of sight is closely aligned to the jet axis ($\theta_{\text{obs}}<10^\circ$).
This results in strongly Doppler boosted non-thermal emission, characterized by the Doppler factor $\delta = \left( \Gamma \, [1 - \beta_{\Gamma} \cos\theta_{\rm obs}] \right)^{-1}$, where $\beta_{\Gamma}$ is the velocity (normalized to the speed of light) corresponding to $\Gamma$. The $\nu F_{\nu}$ peak fluxes are boosted by a factor $\delta^4$ into the observer's frame.
Additionally, radiation frequencies are boosted by a factor $\delta$ and timescales in the co-moving frame are shortened in the observer's frame by a factor $\delta^{-1}$.

Our code solves the time dependent Fokker-Planck and radiation-transfer equations for the lepton and photon populations in the emission region.
The calculations account for radiative cooling from synchrotron emission and inverse Compton scattering of the synchrotron spectrum itself (synchrotron-self Compton; SSC; \cite{konigl_1981,marscher_and_gear_1985,ghisellini_and_maraschi_1989}) as well as external photon fields, i.e. accretion disc, broad-line region and/or dusty torus photons \citep{ghisellini_and_madau_1996,böttcher_1997,blazejowski_2000,ghisellini_and_tavecchio_2008}.
Additionally $\gamma$-$\gamma$ pair production and external photon absorption effects are accounted for in the model.
Accelerated relativistic leptons are injected at each time step with a power law energy spectrum, representing the typical non-thermal electron spectra resulting in most known astrophysical acceleration mechanisms.

\subsection{Blazar Parameters}
\label{sub:blazar_parameters}

Table \ref{tab:blazar_params} shows the parameters for two synthetic blazar types used in this work, similar to the ones used in \cite{thiersen_2022}: a flat spectrum radio quasar (FSRQ) and a high synchrotron-peaked BL Lac object (HBL).
The parameters for these blazars were chosen to be as closely matched as possible while maintaining the characteristics of their classification.
This is done to eliminate unintentional side effects due to differences in blazar parameters as far as possible.
Although observed HBLs generally exhibit redshifts $z \lesssim 0.5$, we use the same redshift for both source types in order to allow for a direct comparison. Placing the simulated HBL at $z=1.0$ implies decreased fluxes and a slight shift of the SED towards lower frquencies, which do not meaningfully impact the variability results.

The differences in parameters are mainly the different lepton spectra of the two blazars.
Compared to the leptons in the FSRQ model, the leptons in HBL models possess higher Lorentz factors to produce the necessary significant SSC emission in X-rays up to VHE $\gamma$-rays.
However, the synchrotron emission in radio up to soft X-rays also produced by HBL-leptons would be overestimated with magnetic fields strengths similar to the FSRQ case, which therefore warrants the lower magnetic field strength for the HBL.
The higher Lorentz factors of the leptons in the HBL case increase the peak frequency of the synchrotron emission such that the spectrum can be classified as HBL.

\begin{figure*}
    \centering
    \includegraphics[width=.48\textwidth]{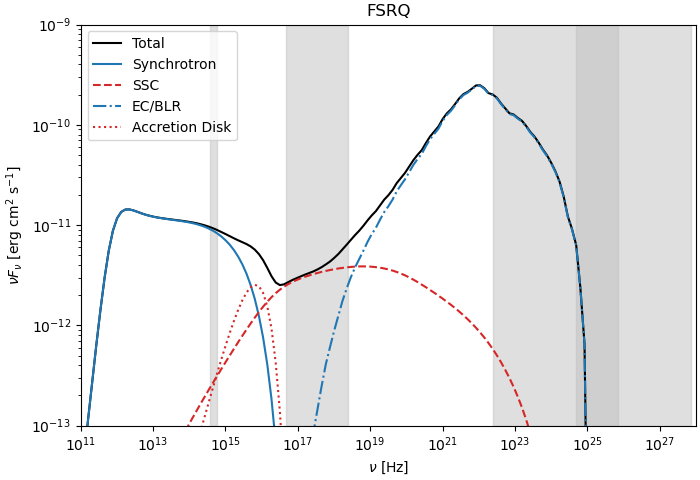}
    \includegraphics[width=.48\textwidth]{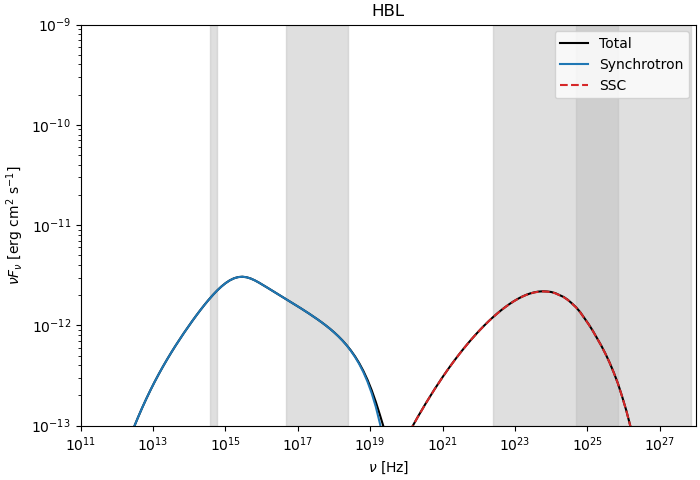} \\
    \caption{
        Steady-state spectral energy distributions (SEDs) for the FSRQ
        (\emph{left}) and HBL (\emph{right}). The gray bands depict the energy
        ranges over which the flux is integrated to generate the optical,
        X-ray, and HE and VHE $\gamma$-ray light curves.
    }
    \label{fig:seds}
\end{figure*}

Furthermore, the differences in electron injection index and co-moving injection luminosity of the electron spectrum are adjusted to balance the spectral energy distributions (SEDs) to be realistic.
Figure \ref{fig:seds} shows the steady-state SEDs of the two blazars, illustrating also which cooling mechanisms dominate in each case.

We derive light curves in the optical ($R$ band), X-ray (0.2 -- 10 keV), HE (100 MeV -- 300 GeV), and VHE (20 GeV -- 300 TeV) $\gamma$-ray domains as indicated in Fig.~\ref{fig:seds}.
These domains are chosen to be representative of current and future instruments such as \textit{Swift}-XRT, \textit{Fermi}-LAT and CTAO.

\begin{deluxetable*}{llcc}
    \tablecaption{
        Initial model parameters for the representative FSRQ and HBL blazar
        cases. These parameters remain fixed unless they represent the varying
        parameter of the respective simulation realization. The subscript $0$
        indicates the parameter's initial value.
    }
    \label{tab:blazar_params}
    \tablewidth{0pt}
    \tablehead{
    \colhead{Definition} & \colhead{Symbol} & \colhead{FSRQ} & \colhead{HBL}
    }
    \startdata
        Magnetic field  & $B_0$ & 1.70 G & 0.40 G \\
        Blob radius & $R$ & $ 3.0 \times 10^{16}$  cm & $3.0 \times 10^{16}$ cm \\
        Ratio of the acceleration to escape time scales & $\eta$ &  1.00 & 1.00  \\
        Escape time scale & $t_{\text{esc}}$  &  $10.0\ R/c$ & $10.0\ R/c$ \\
        Redshift to the source & $z$ &  1.0 &  1.0 \\
        Minimum Lorentz factor of the electron injection spectrum & $\gamma_{\text{min}}$ &  $1.0 \times 10^2$ & $1.0 \times 10^4$\\
        Maximum Lorentz factor of the electron injection spectrum & $\gamma_{\text{max}, 0}$ & $1.0 \times 10^4$ & $1.0 \times 10^6$\\
        Bulk Lorentz factor & $\Gamma$ &  20.0 & 20.0 \\
        Observing angle relative to the axis of the BH jet & $\theta_{\text{obs}}$  &  $5.0 \times 10^{-2}$ rad & $5.0 \times 10^{-2}$ rad\\
        Doppler factor & $\delta$ &  20.0 & 20.0 \\
        Electron injection index & $q_0$ &  2.8 & 2.5\\
        Co-moving injection luminosity  of the electron spectrum  & $L_{\text{inj}, 0}$ &  $5.0 \times 10^{43}$ erg/s & $1.0 \times 10^{42}$ erg/s\\
        Mass of the super massive black hole & $M_{\text{BH}}$ &  $8.5 \times 10^8 M_{\odot}$  & $8.5 \times 10^8 M_{\odot}$ \\
        Eddington ratio & $l_{\text{Edd}}$ &  $1.0 \times 10^{-1}$ & $1.0 \times 10^{-4}$\\
        Initial location of the blob along jet axis  & $d$ &  $6.5 \times 10^{17}$ cm & $6.5 \times 10^{17}$ cm\\
        Radius of the BLR  & $R_{\text{BLR}}$ &  $6.7 \times 10^{17}$ cm & - \\
        Effective temperature of the BLR  & $T_{\text{eff}}$ & $ 5.0 \times 10^{4}$  K & - \\
        Effective luminosity of the BLR  & $L_{\text{BLR}}$ &  $1.0 \times 10^{45}$ erg/s & - \\
    \enddata
\end{deluxetable*}

\subsection{Stochastic Variations}
\label{sub:stochastic_variations}

We assume that a self-sustaining process exists that can change a parameter or parameters in the emission region following our generated stochastic variations.
For simplicity and to isolate the signatures of variations of individual parameters, the stochastic variations are limited to vary only a single parameter in each realization of the model.

Four different parameters were used as varying parameters: the maximum electron Lorentz factor, $\gamma_{\text{max}}$, the electron injection luminosity, $L_{\text{inj}}$, the magnetic field strength, $B$, and the electron spectral index, $q$.
All these parameters are plausible candidates to change in different acceleration processes \citep{summerlin_and_baring_2012,moscibrodzka_2013,zech_and_lemoine_2021,zhang_2022}.

The algorithm for producing synthetic light curves by \cite{timmer_and_koenig_1995} is employed here to produce the variations.
The algorithm takes a power spectral density (PSD) as input to produce a periodogram -- the Fourier transform of a time series -- that maps to a noisy version of the input PSD.
This noise is generated by means of mapping normally distributed random variables that depend on the input PSD.
Applying an inverse Fourier transform on the periodogram produces a time series which represents the synthetic light curve.

The time series is manipulated appropriately to represent a scale factor or an offset of the varying parameters and to fall within physical limits both numerically and physically imposed.
The respective parameter scale factors span the following ranges: $\gamma_{\text{max}}/\gamma_{\text{max,0}}~\in~[0.1,4]$, $L_{\text{inj}}/L_{\text{inj,0}}~\in~[0, 4]$, $B/B_0~\in~[10^{-2}, 3]$, \mbox{$q~\in~[q_0-1,q_0+1]$}.\footnote{Parameters subscripted with $0$ indicate the value at the steady-state SED solution.}
These manipulations are such that the underlying indices in the PSDs are not affected.
Five different values of the PSD index $\alpha$ of the variations were explored: $\alpha = 1.0, 1.5, 2.0, 2.5$, and $3.0$.

\subsection{Variation Permutations}
\label{sub:variation_case}

Variability is generated for each permutation of blazar parameters, variation PSD indices and varying parameters -- 40 permutations in total.
For each permutation a total of 100 simulations were run to obtain good statistics.
Furthermore, in order to test the robustness of our results and ensure that there are no artefacts due to the finite time resolution of the simulations, we run every simulation twice, once with a short time step and once with a long time step, with the respective time steps differing by a factor of $\sim 20$.
In addition  to testing the statistical robustness of our simulations, this also allows us to explore an extended range of temporal frequencies,  referred to as short- and long-time-scale frequencies for the short- and long-time-steps, respectively.

The short-time-step case used a co-moving time step of $\Delta t = 2$ hours which translates to $\Delta t_{\text{obs}} = 720$ seconds in the observer's frame.
Cooling time scales for the radiation mechanisms in the observer's frame given the blazar parameters in Table \ref{tab:blazar_params} range between $\sim 10^2$ seconds up to $\sim 10^5$ seconds depending on the energy of the leptons \citep{thiersen_2022}.
Due to Doppler and redshift effects this time step duration translates to the effective temporal frequencies in the observer's frame falling within the range $[ 7.18\times10^{-7}, 2.9\times10^{-3} ]$~Hz.

Similarly, in the long-time-step case $\Delta t_{\text{obs}} = 1.5\times10^{4}$~s ($\Delta t = 1.5\times 10^5$ seconds in the co-moving frame) which is much longer than most of the cooling time scales of the emission processes involved.
This case probes temporal frequencies in the range $[ 3.48\times10^{-9}, 1.41\times10^{-5}]$~Hz in the observer's frame.
The overlap, $[ 7.18\times10^{-7}, 1.41\times10^{-5}]$~Hz, allows for testing the consistency of the model for different time step sizes.

Light curves were generated for a total of 8000 simulations.
For each of the permutations a waveband-specific average fractional variability \citep{vaughan_2003,schleicher_2019}, and PSD power law index is calculated from the light curves.
The average fractional variability for a permutation is calculated from the fractional variability of each of the light curves of its 100 simulations.
The standard deviation of the fractional variability data is used as error bar.
The waveband-specific average PSD power law index of a permutation is derived by first calculating an average PSD.
This is done by calculating the average and standard deviation for the power in each temporal frequency over the 100 simulations of a permutation.
A power law is then fitted to the average PSD to determine the average PSD index.

\section{Results}
\label{sec:results}

\begin{figure*}[ht!]
    \centering
    \includegraphics[width=.48\textwidth]{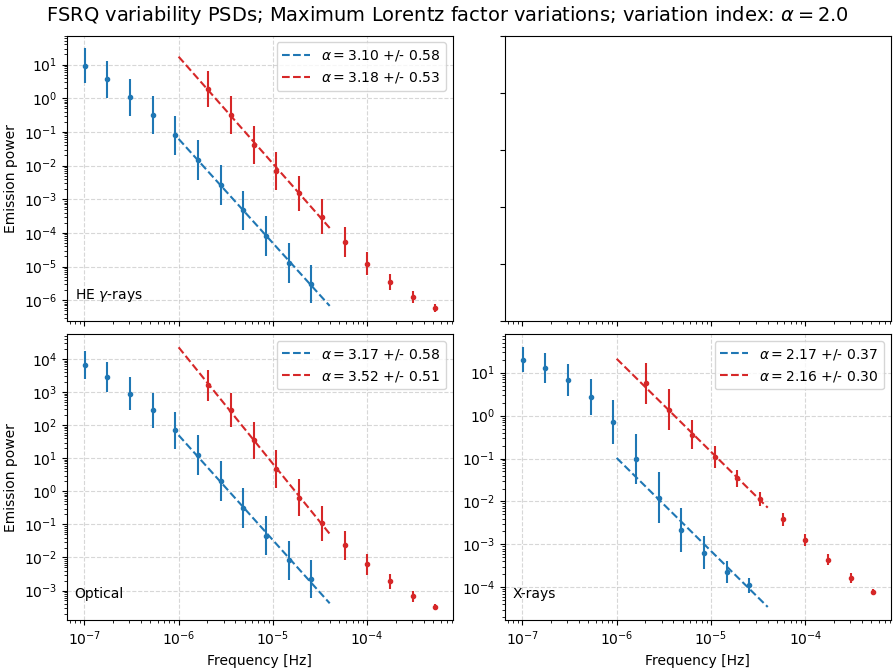}
    \includegraphics[width=.48\textwidth]{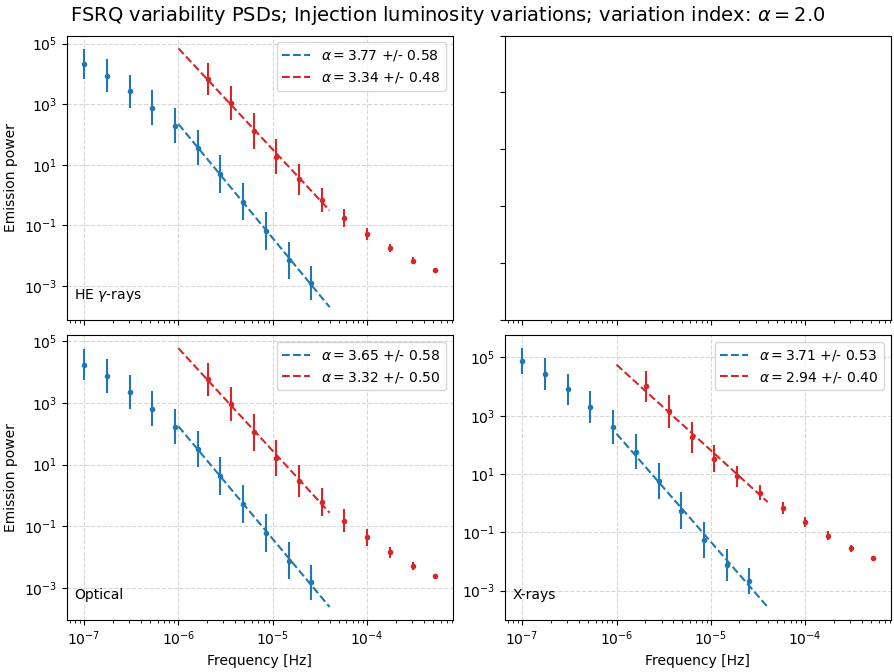} \\
    \includegraphics[width=.48\textwidth]{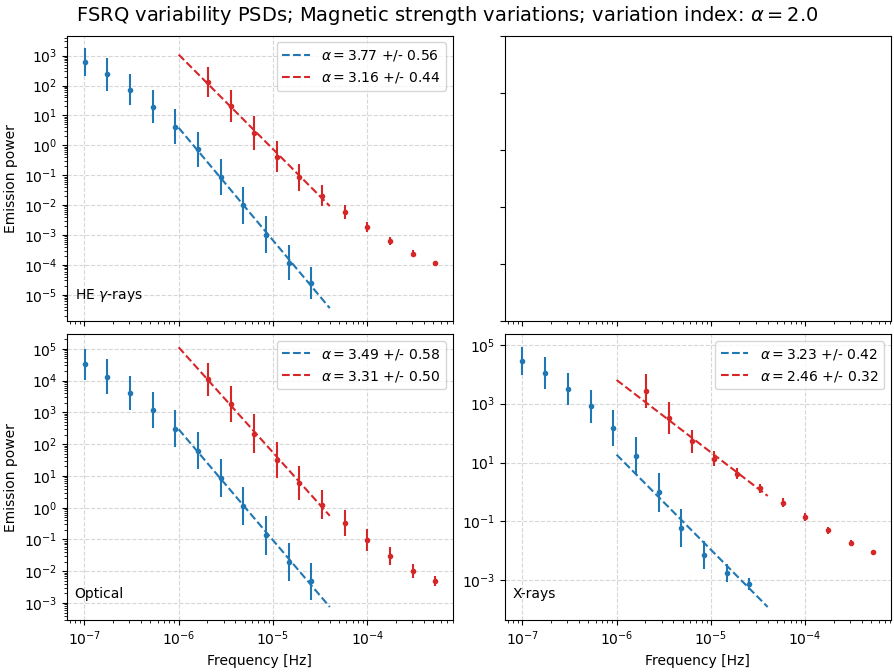}
    \includegraphics[width=.48\textwidth]{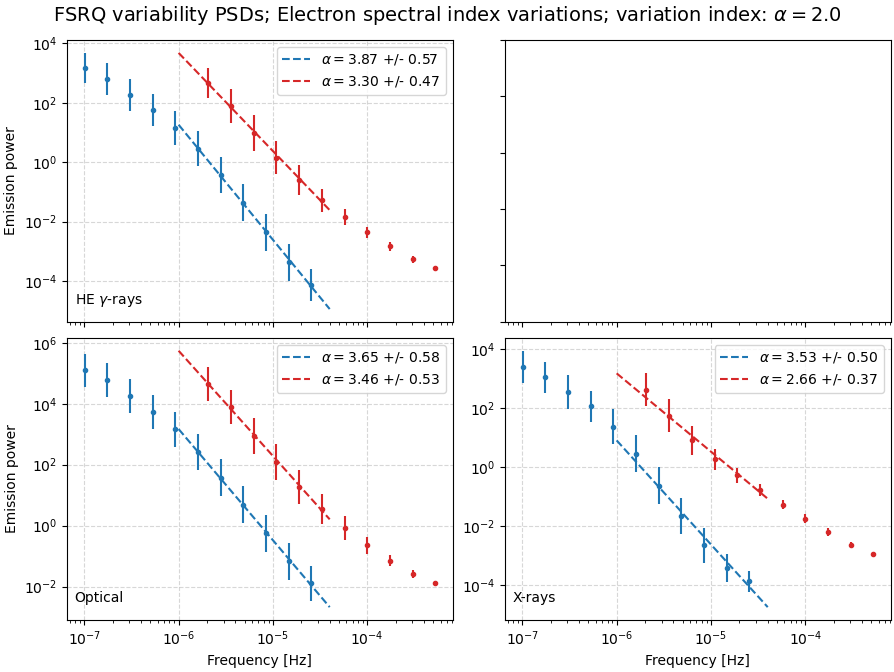}
    \caption{
        Variability PSDs for the FSRQ case with variation power law index
        $2.0$. \emph{Blue}: long-time-scale frequencies. \emph{Red}:
        short-time-scale frequencies. Varying parameters as indicated in the
        titles and wavelengths as indicated in the panels.
    }
    \label{fig:fsrq_200}
\end{figure*}

\begin{figure*}[ht!]
    \includegraphics[width=.48\textwidth]{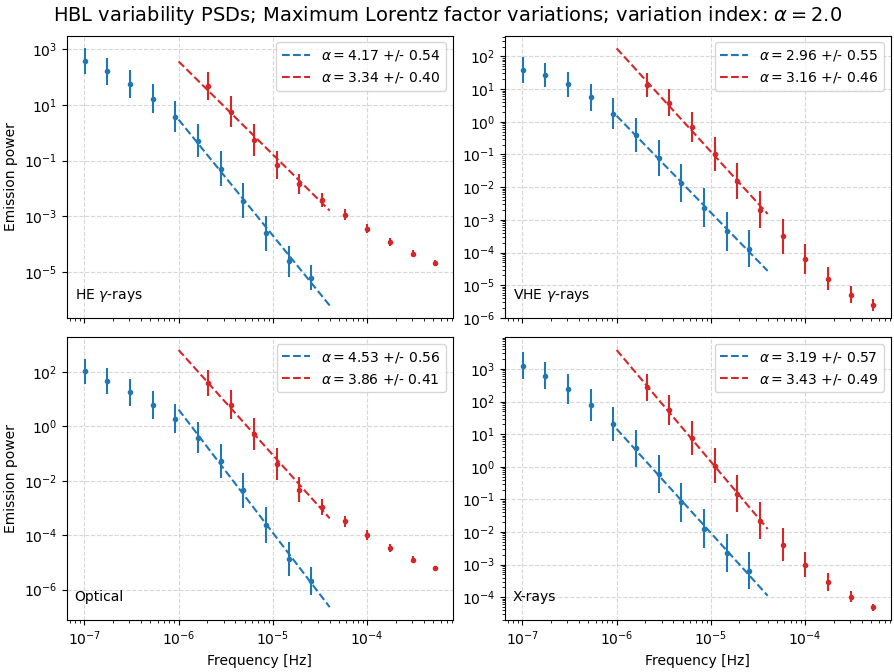}
    \includegraphics[width=.48\textwidth]{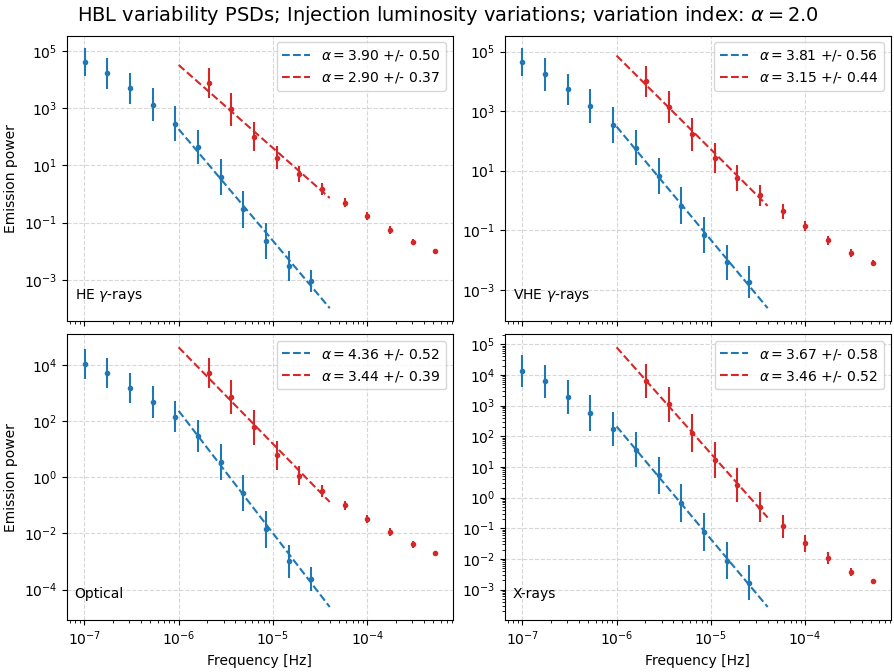} \\
    \includegraphics[width=.48\textwidth]{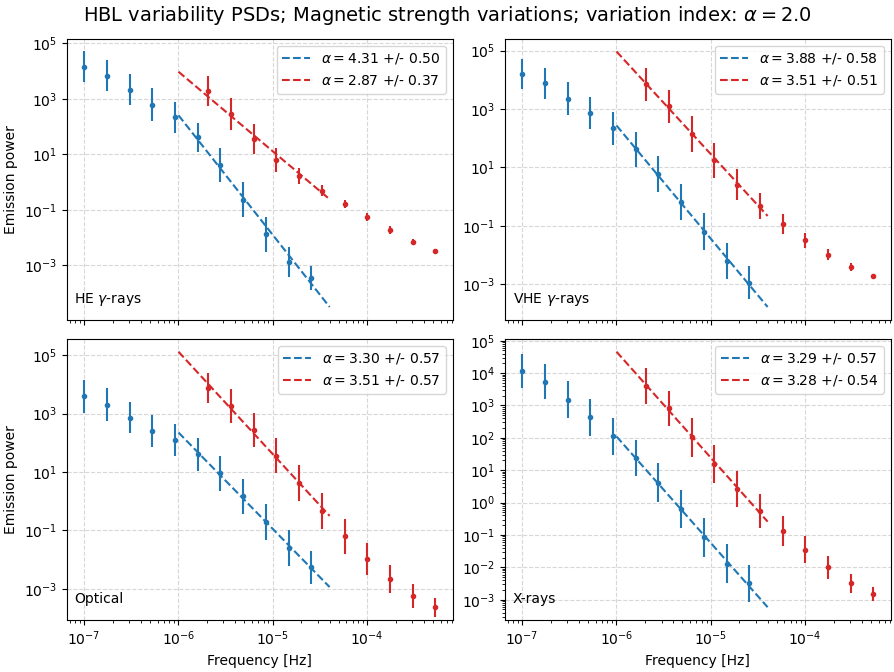}
    \includegraphics[width=.48\textwidth]{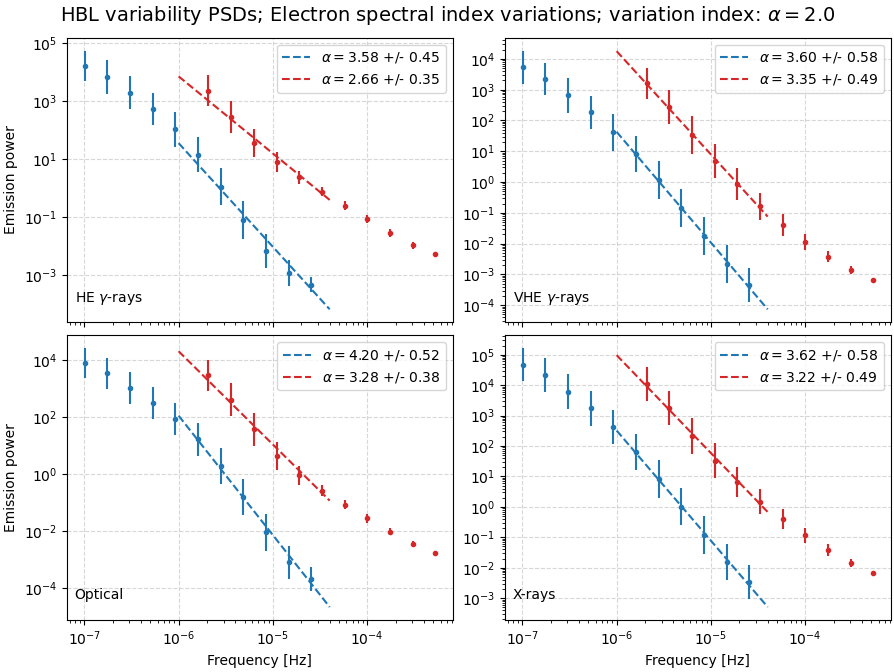}
    \caption{Same as Fig.~\ref{fig:fsrq_200}, but for the HBL case.}
    \label{fig:hbl_200}
\end{figure*}

\begin{figure*}[ht!]
    \centering
    \includegraphics[width=.48\textwidth]{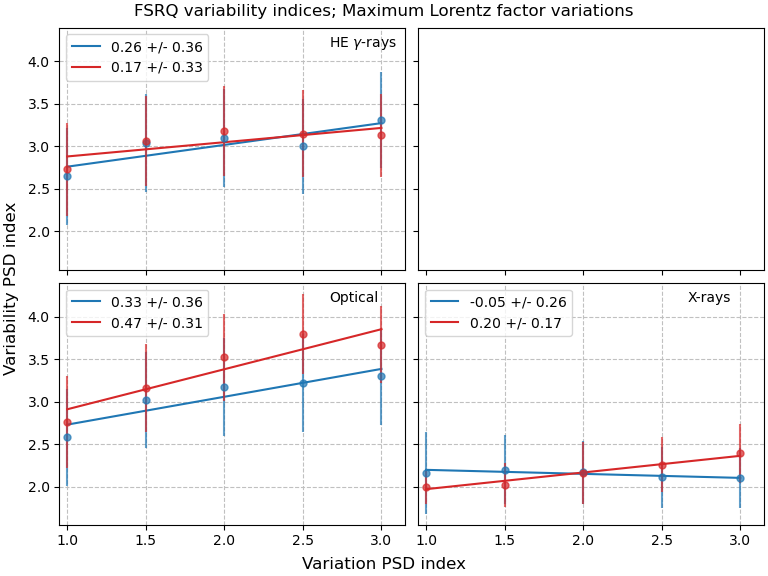}
    \includegraphics[width=.48\textwidth]{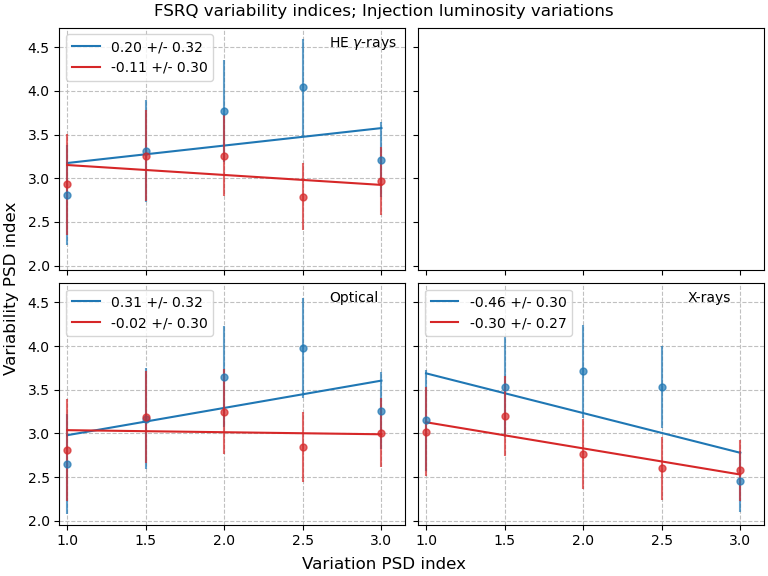} \\
    \includegraphics[width=.48\textwidth]{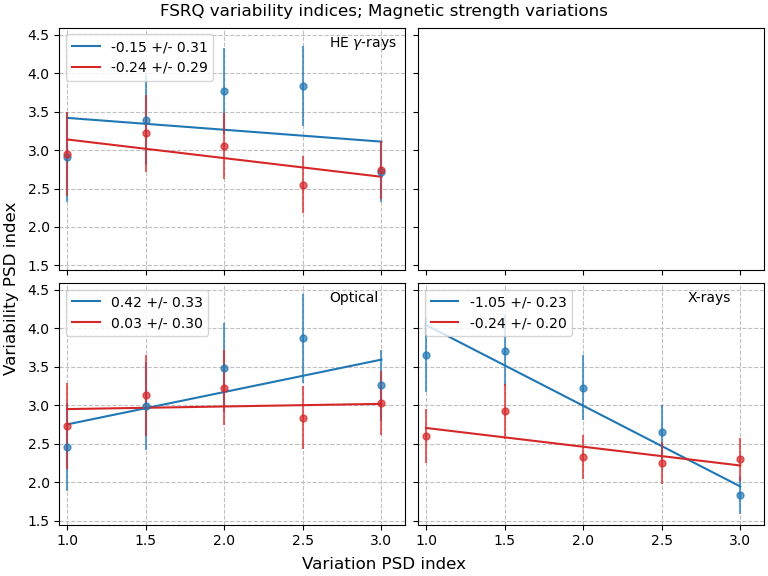}
    \includegraphics[width=.48\textwidth]{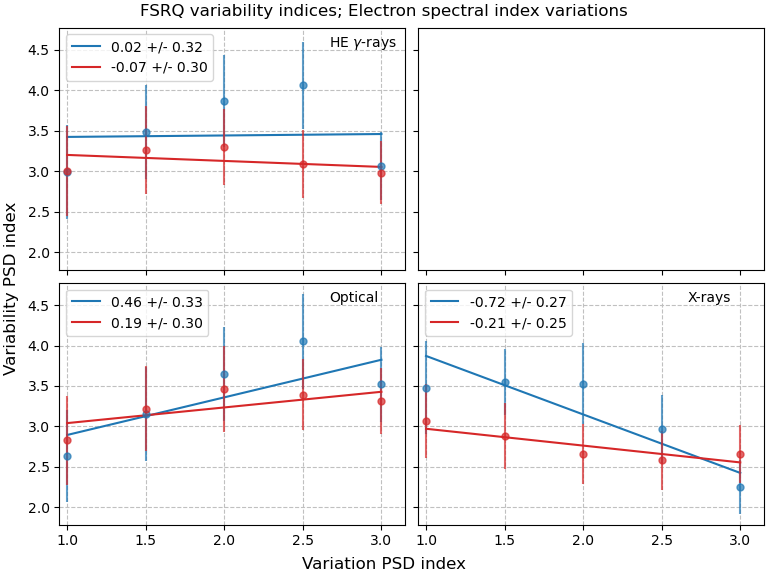}
    \caption{
        PSD index comparisons for the FSRQ simulations. \emph{Blue}:
        long-time-scale frequencies; \emph{Red}: short-time-scale frequencies.
        Varying parameters as indicated in the titles and wavebands as
        indicated in each panel.
    }
    \label{fig:fsrq_comp}
\end{figure*}

\begin{figure*}[ht!]
    \centering
    \includegraphics[width=.48\textwidth]{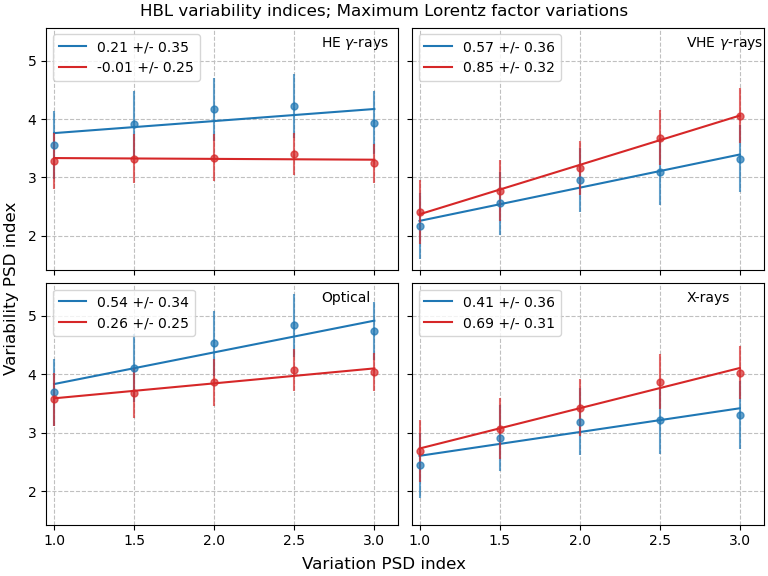}
    \includegraphics[width=.48\textwidth]{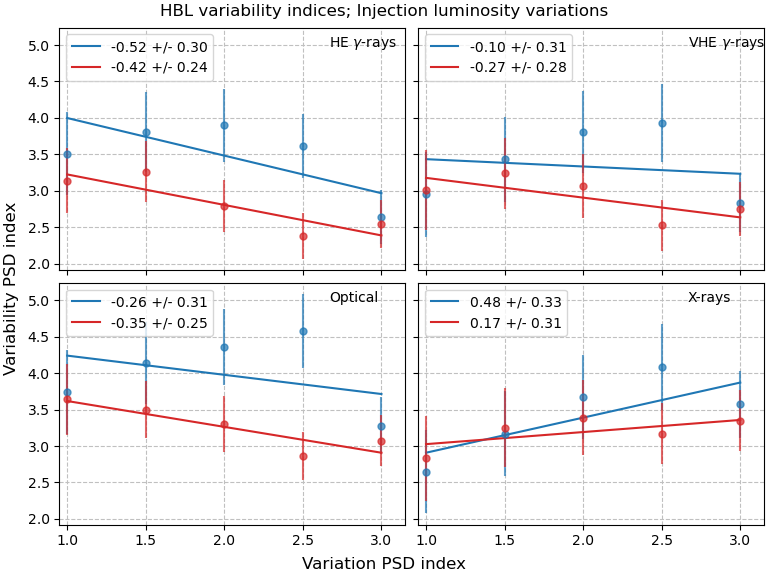} \\
    \includegraphics[width=.48\textwidth]{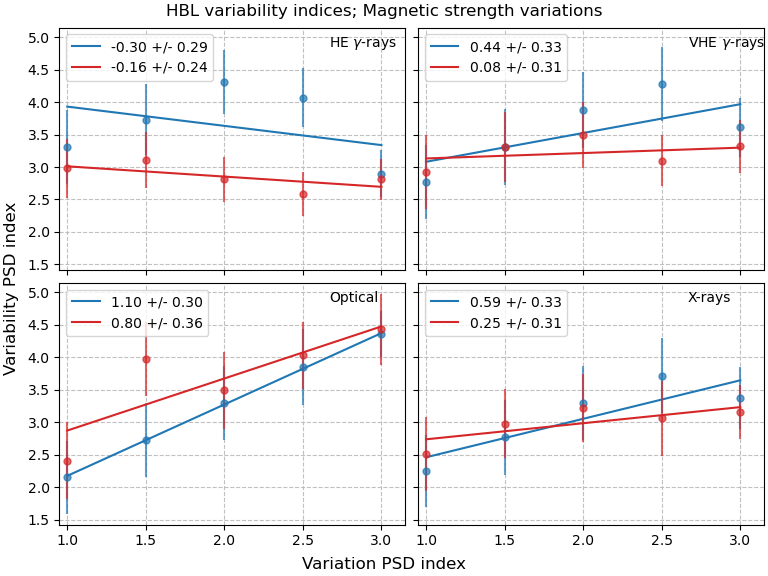}
    \includegraphics[width=.48\textwidth]{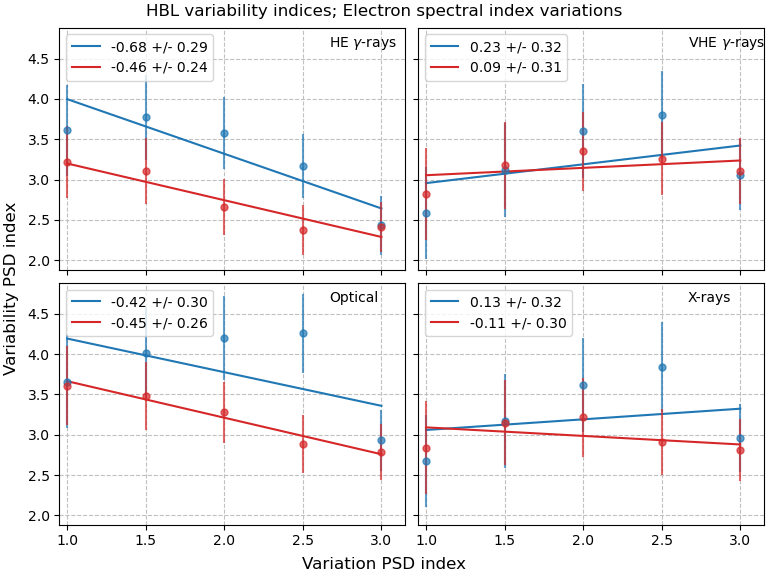}
    \caption{Same as Fig.~\ref{fig:fsrq_comp}, but for the HBL case.
    }
    \label{fig:hbl_comp}
\end{figure*}

\begin{figure*}[ht!]
    \centering
    \includegraphics[width=.48\textwidth]{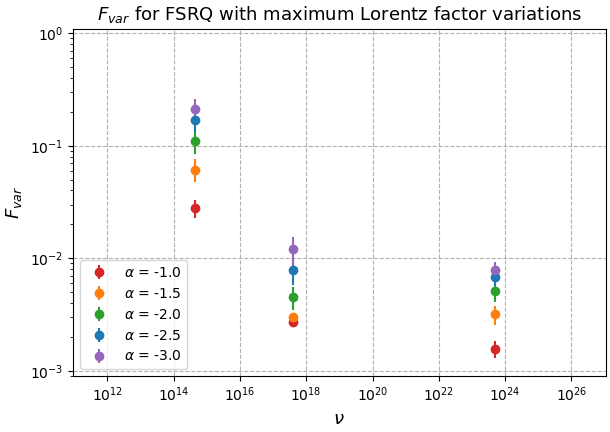}
    \includegraphics[width=.48\textwidth]{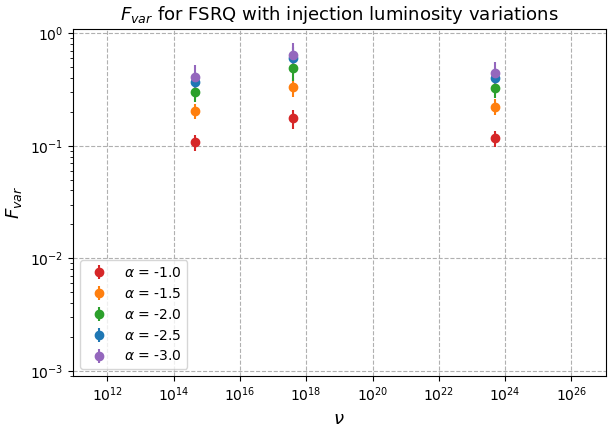} \\
    \includegraphics[width=.48\textwidth]{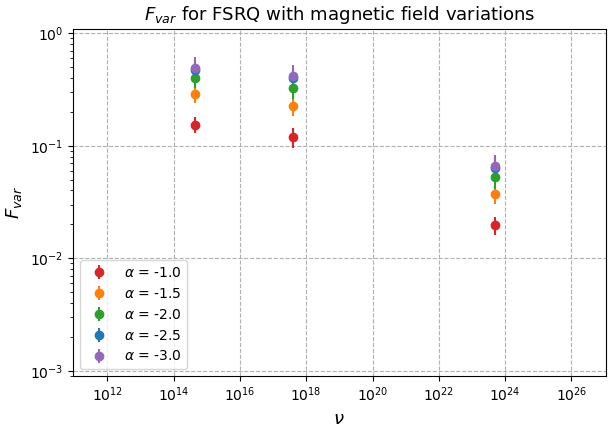}
    \includegraphics[width=.48\textwidth]{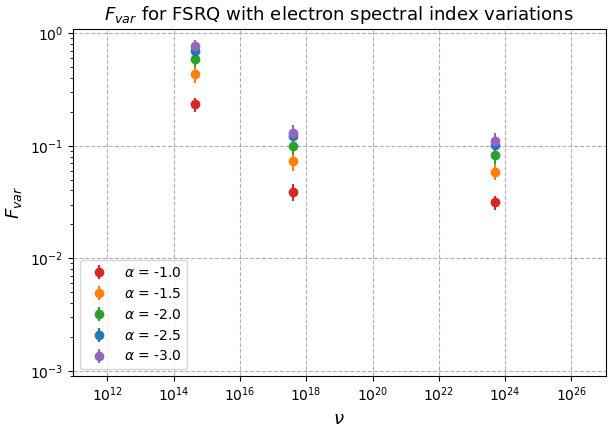}
    \caption{
        Fractional variability for the FSRQ case with colours indicating the
        short-/long-time-step simulations and PSDs fit by power-laws with
        indices provided in the legend.
    }
    \label{fig:fsrq_F_var}
\end{figure*}

\begin{figure*}[ht!]
    \centering
    \includegraphics[width=.48\textwidth]{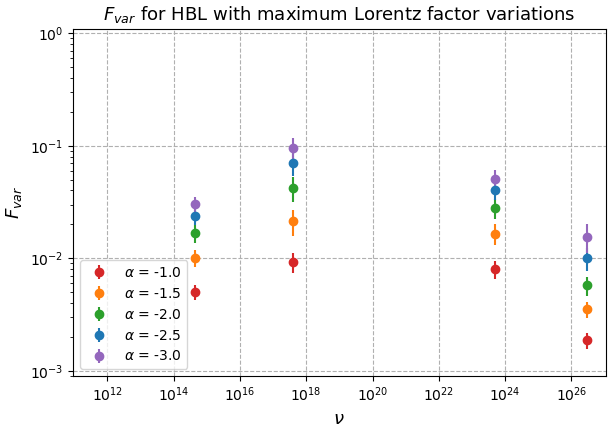}
    \includegraphics[width=.48\textwidth]{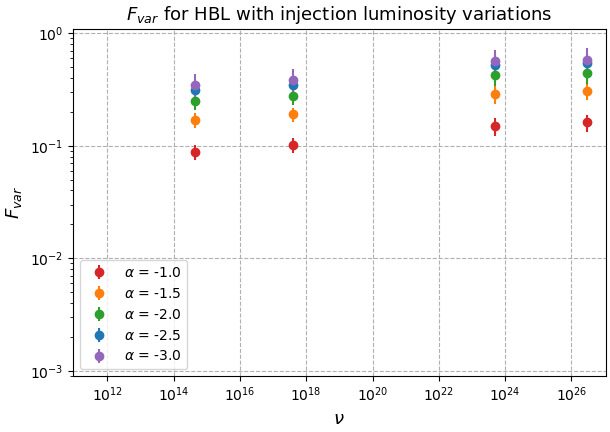} \\
    \includegraphics[width=.48\textwidth]{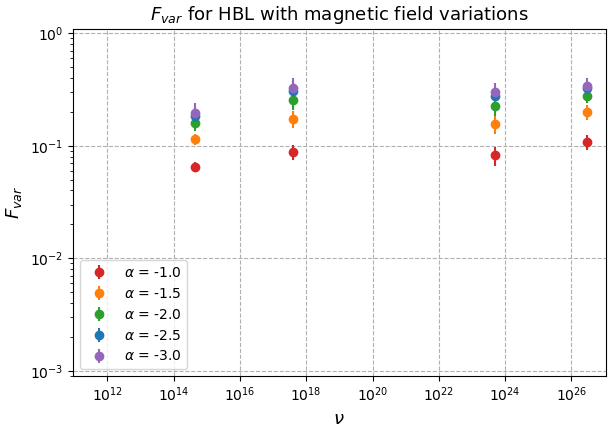}
    \includegraphics[width=.48\textwidth]{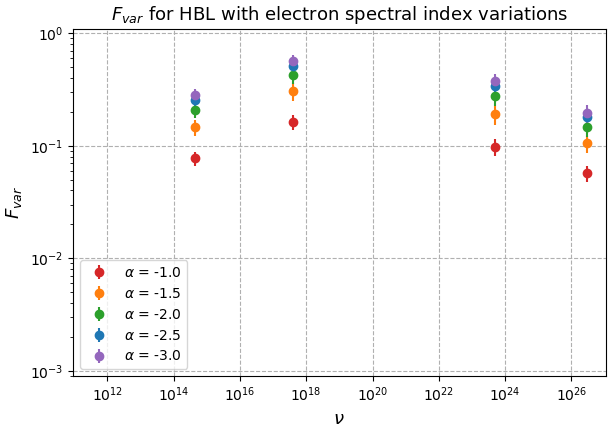}
    \caption{
        Same as Fig.~\ref{fig:fsrq_F_var}, but for the HBL case.
    }
    \label{fig:hbl_F_var}
\end{figure*}

Figures \ref{fig:fsrq_200} and \ref{fig:hbl_200} show the averaged PSDs of simulated variability for each permutation of blazar parameters and varying parameter with input PSD power law index of $2.0$.
The variability indices within the overlapping frequency range are compatible within error margins.
The power differences between the short- and long-time-scale simulations are due to keeping the magnitude of input parameter variations the same for both time step cases.
This leads to a proportional shift in the power of each temporal frequency.
Most notably, the variability indices for all cases are larger than the variation index of $2.0$.

We do not show PSDs for the other variation indices, as they look rather similar to Figs.~\ref{fig:fsrq_200} and \ref{fig:hbl_200}.
The main results of these plots are the variability indices in the overlap region of the short- and long-time-scale simulations.
The comparison is shown in Figs.~\ref{fig:fsrq_comp} and \ref{fig:hbl_comp} for all parameter variations and variation indices.
The results show that the short- and long-time-scale indices are mostly consistent with each other.
With few exceptions, the comparison between the variation and variability PSD indices shows no strong evidence of proportionality as most results are compatible with a constant index of the variability within error margins of $2 \, \sigma$.

The exceptions from the aforementioned compatibility with constant variability index are for the FSRQ the X-ray variability indices for varying (a.1) magnetic field and (a.2) electron spectral index; and for the HBL the (b.1) X-ray and (b.2) VHE $\gamma$-ray variability for varying maximum Lorentz factor, (c) optical variability for magnetic field variations and (d) HE $\gamma$-rays for electron spectral index variations.

In some cases the variability produced in different wavebands in the model is of so small amplitude that it can be considered practically unobservable.
Figures \ref{fig:fsrq_F_var} and \ref{fig:hbl_F_var} show the averaged fractional variability, $F_{\text{var}}$, as a function of frequency for the different permutation cases.
The fractional variability of the simulated light curves is low compared to typical observations, in some cases with values as low as $10^{-3}$.
This is likely due to relatively low amplitudes of the input parameter variations, but is not expected to have a significant impact on the PSD characteristics of the light curves.
As expected, the fractional variability increases with steeper variation temporal indices, as this leads to increased power at low frequencies of the underlying parameter variations.

\section{Discussion and Conclusion}
\label{sec:discussion_and_conclusions}

This work shows that a time-dependent one-zone leptonic blazar model with stochastic parameter variations produces  variability patterns that are almost independent of the PSD power-law index of the input parameter variations and of the time step used for the simulations.
However, it is unphysical for such a power law trend to extend indefinitely to lower temporal frequencies.
Therefore, some cut-off is expected, but no clear indications of such a cut-off were found in our results.

The consistency of a constant temporal spectral index for the variability suggests that it is unlikely that the PSDs of blazar variability are dominated by the temporal characteristics of the underlying physical conditions in blazars, specifically the changes in the injected electron spectrum (i.e. maximum Lorentz factor, injection luminosity, and electron spectral index) and magnetic field variations tested in this work.
As a caveat, we note that the magnetic-field variations simulated in our work do not reflect the drastic changes expected in case of magnetic reconnection.

This leads to the conclusion that the variability  power law index is not intrinsically determined by the temporal spectrum of underlying single parameter fluctuations in the emission region.
However, the temporal variations of physical parameters are still expected to be a plausible cause for small amplitude flux variations which continuously occur even in quiescent states in blazars.

A further caveat is that the fractional variability produced by our simulations is significantly smaller than typically observed in blazar light curves.
While this could be remedied by larger-amplitude parameter variations in our simulations, we do not expect this difference to have a significant impact on the PSD characteristics of the simulated light curves.

In our previous work \citep{thiersen_2022} we studied only a single variation index ($2.0$) and concluded that the resulting PSD index was similar to the variation index. However, a fit of the PSD was not done, and thus the results in both studies are compatible.
However, we concluded in \cite{thiersen_2022} that there is a high likelihood that the variability PSD reflected the PSD of the varying parameter.
This is refuted in this work given that the variability index as a function of the variation index is compatible with a constant value throughout the various cases.
As in our previous work, we find that the model does not produce waveband-dependent variability spectral indices as found in observations.
This is an indication that in reality blazar emission does not originate from a single emission zone; see, e.g., \cite{aharonian_2023} for an example of observational evidence for this. Therefore, a multi-zone emission model with emission zones quasi-causally connected would be better suited for simulation of variability that is more representative of observed variability patterns.
Furthermore, one can also argue that single parameter variations are highly unlikely and not physically plausible which supports the hypothesis of using multi-parameter variations. This is left for future investigations.

A similar study was conducted by \cite{polkas_2021}, but they employed observed light curves from Fermi-LAT in order to generate single parameter variations.
Similar to our conclusions, they found that single parameter variations are unable to reproduce the observed variability.

\begin{acknowledgments}
    The authors acknowledge the following: the Centre for High Performance
    Computing (CHPC) in Cape Town, South Africa, for providing access to
    computational resources and technical support that facilitated the
    simulations presented in this paper.
    MZ acknowledges funding by the Deutsche Forschungsgemeinschaft (DFG, German Research Foundation) -- project number 460248186 (PUNCH4NFDI).
\end{acknowledgments}



\bibliography{sample631}{}
\bibliographystyle{aasjournal}


\end{document}